\pdfoutput=1

\documentclass[12pt,a4paper]{article}

\usepackage{ifthen} 
\newboolean{pdflatex}
\setboolean{pdflatex}{true} 

\newboolean{articletitles}
\setboolean{articletitles}{true} 

\newboolean{uprightparticles}
\setboolean{uprightparticles}{false} 

\newboolean{inbibliography}
\setboolean{inbibliography}{false} 

\def\paperauthors{LHCb collaboration} 
\def\paperasciititle{Updated determination of D0-D0bar mixing and CP violation parameters with D0->K+pi- decays} 
\def\papertitle{Updated determination of \Dz--\Dzb mixing and \CP violation parameters with $\Dz\to K^+\pi^-$ decays} 
\def\paperkeywords{{High Energy Physics}, {LHCb}, {Flavour Physics}, {Charm Physics}, {CP Violation}, {Oscillation}} 
\def\papercopyright{CERN on behalf of the LHCb collaboration}
\def\paperlicence{CC-BY-4.0}
\def\paperlicenceurl{https://creativecommons.org/licenses/by/4.0/}
\def\cernepnumber{CERN-EP-2017-304}
\def\lhcbpapernumber{LHCb-PAPER-2017-046}
\def\paperdate{December 8, 2017}

\newcommand{\pis}{\ensuremath{\pi_{\rm s}}\xspace}
\newcommand{\M}{\ensuremath{M(\Dz\pis^+)}\xspace}
\newcommand{\Ripred}{\ensuremath{\widetilde{R}_i}}

\newcommand{\chisquaredexpression}{
\begin{equation}\label{eqn:fit}
\chi^2 = \sum_i \left[\left(\frac{ r^+_i-\epsilon_r^+\Ripred^+}{\sigma_i^+}\right)^2+
\left(\frac{r^-_i-\epsilon_r^-\Ripred^-}{\sigma_i^-}\right)^2\right]+\chi^2_{\rm corr} \, . 
\end{equation}
}

\usepackage[top=1in, bottom=1.25in, left=1in, right=1in]{geometry}

\usepackage{microtype}
\usepackage{lineno}  
\usepackage{xspace} 
\usepackage{caption} 

\usepackage{graphicx}  
\usepackage{color}
\usepackage{colortbl}
\graphicspath{{./figs/}} 

\usepackage{amsmath} 
\usepackage{amssymb}
\usepackage{amsfonts}
\usepackage{upgreek} 

\newcommand*\patchAmsMathEnvironmentForLineno[1]{%
\expandafter\let\csname old#1\expandafter\endcsname\csname #1\endcsname
\expandafter\let\csname oldend#1\expandafter\endcsname\csname
end#1\endcsname
 \renewenvironment{#1}%
   {\linenomath\csname old#1\endcsname}%
   {\csname oldend#1\endcsname\endlinenomath}%
}
\newcommand*\patchBothAmsMathEnvironmentsForLineno[1]{%
  \patchAmsMathEnvironmentForLineno{#1}%
  \patchAmsMathEnvironmentForLineno{#1*}%
}
\AtBeginDocument{%
\patchBothAmsMathEnvironmentsForLineno{equation}%
\patchBothAmsMathEnvironmentsForLineno{align}%
\patchBothAmsMathEnvironmentsForLineno{flalign}%
\patchBothAmsMathEnvironmentsForLineno{alignat}%
\patchBothAmsMathEnvironmentsForLineno{gather}%
\patchBothAmsMathEnvironmentsForLineno{multline}%
\patchBothAmsMathEnvironmentsForLineno{eqnarray}%
}

\usepackage{hyperxmp}

\usepackage[pdftex,
            pdfauthor={\paperauthors},
            pdftitle={\paperasciititle},
            pdfkeywords={\paperkeywords},
            pdfcopyright={Copyright (C) \papercopyright},
            pdflicenseurl={\paperlicenceurl}]{hyperref}

\usepackage[all]{hypcap} 

\usepackage{xspace} 
\usepackage{upgreek}

\def\lhcb {\mbox{LHCb}\xspace}

\def\MagUp {\mbox{\em Mag\kern -0.05em Up}\xspace}

\ifthenelse{\boolean{uprightparticles}}%
{

 \def\PDelta      {\ensuremath{\Delta}\xspace}                 
 \def\PXi      {\ensuremath{\Xi}\xspace}                 
 \def\PLambda      {\ensuremath{\Lambda}\xspace}                 
 \def\PSigma      {\ensuremath{\Sigma}\xspace}                 
 \def\POmega      {\ensuremath{\Omega}\xspace}                 
 \def\PUpsilon      {\ensuremath{\Upsilon}\xspace}                 
 
 \def\PB      {\ensuremath{\mathrm{B}}\xspace}                 
                  
 \def\PD      {\ensuremath{\mathrm{D}}\xspace}

 \def\PK      {\ensuremath{\mathrm{K}}\xspace}

 \def\Pb      {\ensuremath{\mathrm{b}}\xspace}                 
 \def\Pc      {\ensuremath{\mathrm{c}}\xspace}

 \def\Pi      {\ensuremath{\mathrm{i}}\xspace}

}
{

 \mathchardef\PDelta="7101
 \mathchardef\PXi="7104
 \mathchardef\PLambda="7103
 \mathchardef\PSigma="7106
 \mathchardef\POmega="710A
 \mathchardef\PUpsilon="7107
                  
 \def\PB      {\ensuremath{B}\xspace}                 
                  
 \def\PD      {\ensuremath{D}\xspace}

 \def\PK      {\ensuremath{K}\xspace}

 \def\Pb      {\ensuremath{b}\xspace}                 
 \def\Pc      {\ensuremath{c}\xspace}

 \def\Pi      {\ensuremath{i}\xspace}

}

\makeatletter
\ifcase \@ptsize \relax
  \newcommand{\miniscule}{\@setfontsize\miniscule{4}{5}}
\or
  \newcommand{\miniscule}{\@setfontsize\miniscule{5}{6}}
\or
  \newcommand{\miniscule}{\@setfontsize\miniscule{5}{6}}
\fi
\makeatother

\DeclareRobustCommand{\optbar}[1]{\shortstack{{\miniscule (\rule[.5ex]{1.25em}{.18mm})}
  \\ [-.7ex] $#1$}}

\def\cquark    {{\ensuremath{\Pc}}\xspace}

\def\bquark    {{\ensuremath{\Pb}}\xspace}

\def\kaon    {{\ensuremath{\PK}}\xspace}
  \def\Kbar    {{\kern 0.2em\overline{\kern -0.2em \PK}{}}\xspace}

\def\KorKbar    {\kern 0.18em\optbar{\kern -0.18em K}{}\xspace}

\def\Kp      {{\ensuremath{\kaon^+}}\xspace}
\def\Km      {{\ensuremath{\kaon^-}}\xspace}

\def\KS      {{\ensuremath{\kaon^0_{\mathrm{ \scriptscriptstyle S}}}}\xspace}

  \def\Dbar    {{\kern 0.2em\overline{\kern -0.2em \PD}{}}\xspace}
\def\D       {{\ensuremath{\PD}}\xspace}

\def\DorDbar    {\kern 0.18em\optbar{\kern -0.18em D}{}\xspace}
\def\Dz      {{\ensuremath{\D^0}}\xspace}
\def\Dzb     {{\ensuremath{\Dbar{}^0}}\xspace}
\def\Dp      {{\ensuremath{\D^+}}\xspace}
\def\Dm      {{\ensuremath{\D^-}}\xspace}

\def\Dstar   {{\ensuremath{\D^*}}\xspace}

\def\Dstarp  {{\ensuremath{\D^{*+}}}\xspace}
\def\Dstarm  {{\ensuremath{\D^{*-}}}\xspace}

\def\Bbar    {{\ensuremath{\kern 0.18em\overline{\kern -0.18em \PB}{}}}\xspace}

\def\BorBbar    {\kern 0.18em\optbar{\kern -0.18em B}{}\xspace}

  \def\Y#1S{\ensuremath{\PUpsilon{(#1S)}}\xspace}

\def\Lbar        {{\ensuremath{\kern 0.1em\overline{\kern -0.1em\PLambda}}}\xspace}
\def\LorLbar    {\kern 0.18em\optbar{\kern -0.18em \PLambda}{}\xspace}

\def\to                 {\ensuremath{\rightarrow}\xspace}

\def\CP                {{\ensuremath{C\!P}}\xspace}

\def\AT#1     {\ensuremath{A_{\mathrm{T}}^{#1}}\xspace}           

\def\C#1      {\ensuremath{\mathcal{C}_{#1}}\xspace}                       
\def\Cp#1     {\ensuremath{\mathcal{C}_{#1}^{'}}\xspace}                    
\def\Ceff#1   {\ensuremath{\mathcal{C}_{#1}^{\mathrm{(eff)}}}\xspace}        
\def\Cpeff#1  {\ensuremath{\mathcal{C}_{#1}^{'\mathrm{(eff)}}}\xspace}       
\def\Ope#1    {\ensuremath{\mathcal{O}_{#1}}\xspace}                       
\def\Opep#1   {\ensuremath{\mathcal{O}_{#1}^{'}}\xspace}                    

\newcommand{\tev}{\ifthenelse{\boolean{inbibliography}}{\ensuremath{~T\kern -0.05em eV}}{\ensuremath{\mathrm{\,Te\kern -0.1em V}}}\xspace}
\newcommand{\gev}{\ensuremath{\mathrm{\,Ge\kern -0.1em V}}\xspace}
\newcommand{\mev}{\ensuremath{\mathrm{\,Me\kern -0.1em V}}\xspace}
\newcommand{\kev}{\ensuremath{\mathrm{\,ke\kern -0.1em V}}\xspace}
\newcommand{\ev}{\ensuremath{\mathrm{\,e\kern -0.1em V}}\xspace}
\newcommand{\gevc}{\ensuremath{{\mathrm{\,Ge\kern -0.1em V\!/}c}}\xspace}
\newcommand{\mevc}{\ensuremath{{\mathrm{\,Me\kern -0.1em V\!/}c}}\xspace}
\newcommand{\gevcc}{\ensuremath{{\mathrm{\,Ge\kern -0.1em V\!/}c^2}}\xspace}
\newcommand{\gevgevcccc}{\ensuremath{{\mathrm{\,Ge\kern -0.1em V^2\!/}c^4}}\xspace}
\newcommand{\mevcc}{\ensuremath{{\mathrm{\,Me\kern -0.1em V\!/}c^2}}\xspace}

\def\mum  {\ensuremath{{\,\upmu\mathrm{m}}}\xspace}

\def\invfb   {\ensuremath{\mbox{\,fb}^{-1}}\xspace}

\def\gsim{{~\raise.15em\hbox{$>$}\kern-.85em
          \lower.35em\hbox{$\sim$}~}\xspace}
\def\lsim{{~\raise.15em\hbox{$<$}\kern-.85em
          \lower.35em\hbox{$\sim$}~}\xspace}

\def\ptot       {\mbox{$p$}\xspace}
\def\pt         {\mbox{$p_{\mathrm{ T}}$}\xspace}

\def\tell1  {TELL1\xspace}
\def\ukl1   {UKL1\xspace}

\usepackage{cite} 
\usepackage{mciteplus}

\usepackage{longtable} 

\begin{document}
\renewcommand{\thefootnote}{\fnsymbol{footnote}}
\setcounter{footnote}{1}


\begin{titlepage}
\pagenumbering{roman}

\vspace*{-1.5cm}
\centerline{\large EUROPEAN ORGANIZATION FOR NUCLEAR RESEARCH (CERN)}
\vspace*{1.5cm}
\noindent
\begin{tabular*}{\linewidth}{lc@{\extracolsep{\fill}}r@{\extracolsep{0pt}}}
\ifthenelse{\boolean{pdflatex}}
{\vspace*{-1.5cm}\mbox{\!\!\!\includegraphics[width=.14\textwidth]{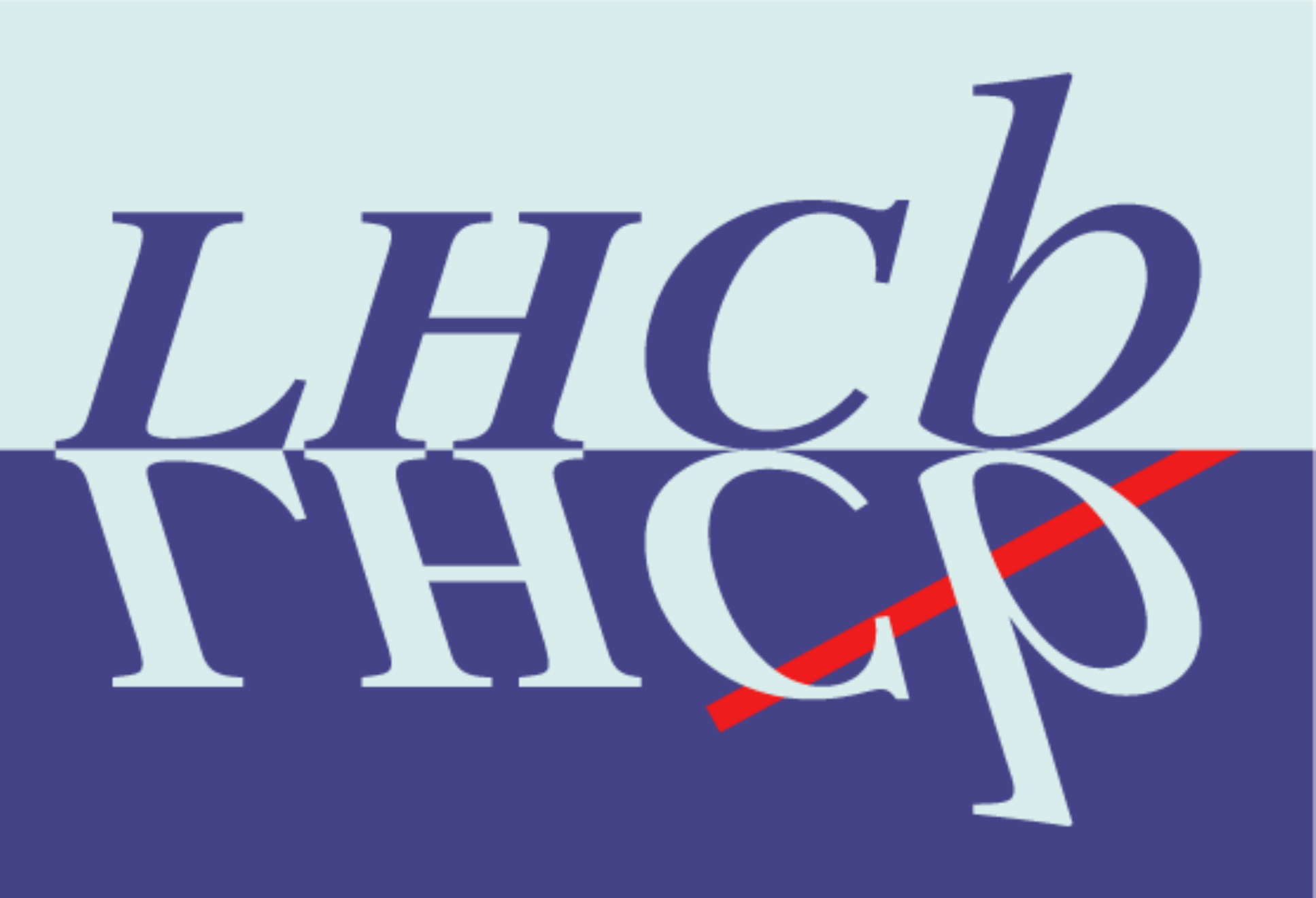}} & &}%
{\vspace*{-1.2cm}\mbox{\!\!\!\includegraphics[width=.12\textwidth]{lhcb-logo.eps}} & &}%
\\
 & & \cernepnumber \\  
 & & \lhcbpapernumber \\  
 & & \paperdate \\ 
 & & \\
\end{tabular*}

\vspace*{3.0cm}

{\normalfont\bfseries\boldmath\huge
\begin{center}
  \papertitle 
\end{center}
}

\vspace*{2.0cm}

\begin{center}
\paperauthors\footnote{Authors are listed at the end of this paper.}
\end{center}

\vspace{\fill}

\begin{abstract}
\noindent 
We report measurements of charm-mixing parameters based on the decay-time-dependent ratio of $\Dz\to K^+\pi^-$ to $\Dz\to K^-\pi^+$ rates. The analysis uses a data sample of proton-proton collisions corresponding to an integrated luminosity of $5.0\invfb$ recorded by the \lhcb experiment from 2011 through 2016. Assuming charge-parity (\CP) symmetry, the mixing parameters are determined to be $x'^2~=~(3.9 \pm 2.7) \times10^{-5}$, $y'~=~(5.28 \pm 0.52) \times 10^{-3}$, and $R_D~=~(3.454 \pm 0.031)\times10^{-3}$. Without this assumption, the measurement is performed separately for \Dz and \Dzb mesons, yielding a direct \CP-violating asymmetry $A_D =(-0.1\pm9.1)\times10^{-3}$, and magnitude of the ratio of mixing parameters $1.00< |q/p| <1.35$ at the $68.3\%$ confidence level.  All results include statistical and systematic uncertainties and improve significantly upon previous single-measurement determinations. No evidence for \CP violation in charm mixing is observed.
\end{abstract}

\vspace*{2.0cm}

\begin{center}
Published in Phys.\ Rev.\  D {\bf 97} (2018) 031101.
\end{center}

\vspace{\fill}

{\footnotesize 
\centerline{\copyright~\papercopyright, licence \href{\paperlicenceurl}{\paperlicence}.}}
\vspace*{2mm}

\end{titlepage}


\newpage
\setcounter{page}{2}
\mbox{~}
%
%
%
%

\cleardoublepage


\renewcommand{\thefootnote}{\arabic{footnote}}
\setcounter{footnote}{0}



\pagestyle{plain} 
\setcounter{page}{1}
\pagenumbering{arabic}


\section{Introduction}
The mass eigenstates of neutral charm mesons are linear combinations of the flavor eigenstates, $|D_{1, 2} \rangle = p | \Dz \rangle \pm  q|\Dzb \rangle$, where $p$ and $q$ are complex-valued coefficients. This results in \Dz--\Dzb  oscillations. In the limit of charge-parity (\CP) symmetry, oscillations are characterized by the dimensionless differences in mass, $x\equiv \Delta m/\Gamma \equiv (m_2 - m_1)/\Gamma$,  and decay width, $y \equiv \Delta \Gamma/2\Gamma \equiv (\Gamma_2 - \Gamma_1)/2\Gamma$, between the \CP-even ($D_2$) and \CP-odd ($D_1$) mass eigenstates,  where $\Gamma$ is the average decay width of neutral \D mesons. If \CP symmetry does not hold, the oscillation probabilities for mesons produced as \Dz and \Dzb can differ, further enriching the phenomenology. Long- and short-distance amplitudes govern the oscillations of neutral \D mesons~\cite{Bianco:2003vb,Burdman:2003rs,Artuso:2008vf}. Long-distance amplitudes depend on the exchange of low-energy gluons and are challenging to calculate. Short-distance amplitudes may include contributions from a broad class of particles not described in the standard model, which might affect the oscillation rate or introduce a difference between the \Dz and \Dzb meson decay rates. The study of \CP violation in \Dz oscillations therefore offers sensitivity to non-standard-model phenomena~\cite{Blaylock:1995ay,Petrov:2006nc,Golowich:2007ka,Ciuchini:2007cw}.\par
The first evidence for \Dz--\Dzb oscillations was reported in 2007~\cite{Aubert:2007wf,Staric:2007dt}. More recently, precise results from the LHCb collaboration~\cite{Aaij:2011ad,Aaij:2012nva, Aaij:2013wda,Aaij:2015xoa,Aaij:2016rhq,Aaij:2016roz} improved the knowledge of the mixing parameters, $x =(4.6_{-1.5}^{+1.4}) \times 10^{-3}$ and $y = (6.2 \pm 0.8)\times 10^{-3}$~\cite{Amhis:2016xyh},  although neither a nonzero value for the mass difference nor a departure from \CP symmetry have been established.
\par
This paper reports measurements of \CP-averaged and \CP-violating mixing parameters in \Dz--\Dzb  oscillations based on the comparison of the decay-time-dependent ratio of $\Dz\to K^+\pi^-$ to $\Dz\to K^-\pi^+$ rates with the corresponding ratio for the charge-conjugate processes. The analysis uses data corresponding to an integrated luminosity of $5.0\invfb$ from proton-proton ($pp$) collisions at 7, 8, and $13\,\tev$ center-of-mass energies, recorded with the \lhcb experiment from 2011 through 2016. This analysis improves upon a previous measurement~\cite{Aaij:2013wda}, owing to the tripling of the sample size and an improved treatment of systematic uncertainties.  The inclusion of charge-conjugate processes is implicitly assumed unless stated otherwise.
\par 
The neutral \D-meson flavor at production is determined from the charge of the low-momentum pion (soft pion), $\pis^{+}$, produced in the flavor-conserving strong-interaction decay $D^{*}(2010)^+\to\Dz\pis^+$. The shorthand notation \Dstarp is used to indicate the $D^{*}(2010)^+$ meson throughout.  We denote as right-sign (RS) the $\Dstarp\to\Dz(\to K^-\pi^+)\pis^+$ process, which is dominated by a Cabibbo-favored amplitude. Wrong-sign (WS) decays, $\Dstarp\to\Dz(\to K^+\pi^-)\pis^+$, arise from the doubly Cabibbo-suppressed $\Dz\to K^+\pi^-$ decay and the Cabibbo-favored $\Dzb\to K^+\pi^-$ decay that follows \Dz--\Dzb oscillation. Since the mixing parameters are small, $|x|,|y|\ll1$, the \CP-averaged decay-time-dependent ratio of WS-to-RS rates is approximated as~\cite{Bianco:2003vb,Burdman:2003rs,Artuso:2008vf,Blaylock:1995ay}
\begin{equation}\label{eq:true-ratio}
R(t) \approx R_D+\sqrt{R_D}\ y'\ \frac{t}{\tau}+\frac{x'^2+y'^2}{4}\left(\frac{t}{\tau}\right)^2,
\end{equation}
where $t$ is the proper decay time, $\tau$ is the average \Dz lifetime, and $R_D$ is the ratio of suppressed-to-favored decay rates. The parameters $x'$ and $y'$ depend on the mixing parameters, $x' \equiv x\cos\delta+y\sin\delta$ and $y' \equiv y\cos\delta-x\sin\delta$, through the strong-phase difference  $\delta$ between the suppressed and favored amplitudes, $\mathcal{A}(\Dz\to K^+\pi^-)/\mathcal{A}(\Dzb\to K^+\pi^-) = -\sqrt{R_D} e^{-i\delta}$, which was measured at the CLEO-c and BESIII experiments~\cite{Asner:2012xb,Ablikim:2014gvw}. If \CP violation occurs, the decay-rate ratios $R^+(t)$ and $R^-(t)$ of mesons produced as \Dz and \Dzb, respectively, are functions of independent sets of mixing parameters, $R_D^\pm,\, (x'^{\pm})^2,\,$ and $y'^\pm$. The parameters $R^+_D$ and $R^-_D$ differ if the ratio between the suppressed and favored decay amplitudes is not \CP symmetric, indicating direct \CP violation. Violation of \CP symmetry either in mixing, $|q/p|\neq1$, or in the interference between mixing and decay amplitudes, $\phi\equiv\arg\left[q\mathcal{A}(\Dzb\to K^+\pi^-)/p\mathcal{A}(\Dz\to K^+\pi^-)\right] \neq \delta$, are referred to as manifestations of indirect \CP violation and generate differences between $((x'^{+})^2,\, y'^+)$ and $((x'^{-})^2,\, y'^-)$.\par
Experimental effects such as differing efficiencies for reconstructing WS and RS decays may bias the observed ratios of signal decays and, therefore, the mixing-parameter results. We assume that the efficiency for reconstructing and selecting the $K^\mp\pi^\pm\pis^+$ final state approximates as the product of the efficiency for the $K^\mp\pi^\pm$ pair from the \Dz decay and the efficiency for the soft pion. The observed
WS-to-RS yield ratio then equals $R(t)$ multiplied by the ratio of the efficiencies for reconstructing $K^+\pi^-$ and $K^-\pi^+$ pairs, which is the only relevant instrumental nuisance. The asymmetry in production rates between \Dstarp and \Dstarm mesons in the LHCb acceptance and asymmetries in detecting soft pions of different charges cancel in the WS-to-RS ratio.

Candidate \Dstarp mesons produced directly in the collision (primary \Dstarp) are reconstructed while suppressing background contributions from charm mesons produced in the decay of bottom hadrons (secondary \Dstarp) and misreconstructed decays. Residual contaminations from such backgrounds are measured using control regions. The asymmetry in $K^\pm\pi^\mp$ reconstruction efficiency is estimated using control samples of charged $D$-meson decays. The yields of RS and WS primary \Dstarp candidates are determined, separately for each flavor, in intervals (bins) of decay time by fitting the \Dstarp mass distribution of candidates consistent with being \Dz decays. We fit the resulting WS-to-RS yield ratios  as a function of decay time to measure the mixing and \CP-violation parameters, including the effects of instrumental asymmetries, residual background contamination, and all considered systematic contributions. To ensure unbiased results, the differences in the decay-time dependence of the WS \Dz and \Dzb samples are not examined until the analysis procedure is finalized. 

\section{The LHCb detector}
The \lhcb detector~\cite{Alves:2008zz} is a single-arm forward spectrometer covering the \mbox{pseudorapidity} range $2<\eta <5$, designed for the study of particles containing \bquark or \cquark quarks. The detector achieves high precision charged-particle tracking using a silicon-strip vertex detector surrounding the $pp$ interaction region, a large-area silicon-strip detector located upstream of a dipole magnet with a bending power of about $4{\mathrm{\,Tm}}$, and three layers of silicon-strip detectors and straw drift tubes placed downstream of the magnet. The tracking system provides a measurement of charged-particle momentum \ptot with a relative uncertainty varying from 0.5\% at low momentum to 1.0\% at 200\gevc. The typical decay-time resolution for $\Dz \to K^+\pi^-$ decays is 10\% of the \Dz lifetime. The polarity of the dipole magnet is reversed periodically throughout data-taking. The minimum distance of a charged-particle trajectory (track) to a proton-proton interaction space-point (primary vertex), the impact parameter,  is measured with $(15+29/\pt)\mum$ resolution, where \pt is the component of the momentum transverse to the beam, in\,\gevc. Charged hadrons are identified using two ring-imaging Cherenkov detectors.  Photons, electrons and hadrons are identified by scintillating-pad and preshower detectors, and an electromagnetic and a hadronic calorimeter. Muons are identified by alternating layers of iron and multiwire proportional chambers. The online event selection is performed by a hardware trigger, based on information from the calorimeter and muon detectors, followed by a software trigger, based on information on displaced charged particles reconstructed in the event. Offline-like quality detector alignment and calibrations, performed between the hardware and software stages, are available to the software trigger for the 2015 and 2016 data~\cite{Dujany:2015lxd,Aaij:2016rxn}. Hence, for these data the analysis uses candidates reconstructed in the software trigger to reduce event size.

\section{Event selection and candidate reconstruction}
Events enriched in  \Dstarp candidates originating from the primary vertex are selected by the hardware trigger by imposing that either one or more \Dz decay products are consistent with depositing a large transverse energy in the calorimeter or that an accept decision is taken independently of the \Dz decay products and soft pion. In the software trigger, one or more \Dz decay products are required to be inconsistent with charged particles originating from the primary vertex and, for 2015 and 2016 data, loose particle-identification criteria are imposed on these final-state particles. Each \Dz candidate is then combined with a low-momentum positive-charge particle originating from the primary vertex to form a \Dstarp candidate. \par In the offline analysis, criteria on track and primary-vertex quality are imposed. To suppress the contamination from misidentified two-body \Dz decays, the pion and kaon candidates from the \Dz decay are subjected to stringent particle-identification criteria. An especially harmful background is generated by a 3\% contribution of soft pions misreconstructed by combining their track segments in the vertex detector with unrelated segments in the downstream tracking detectors. The track segments in the vertex detector are genuine, resulting in properly measured opening angles in the $\Dstarp\to \Dz\pis^+$ decay. Since the opening angle dominates over the $\pis^+$ momentum in the determination of the \Dstarp mass, such spurious soft pions tend to produce a signal-like peak in the \Dstarp mass spectrum. In addition, they bias the WS-to-RS ratio because the mistaken association with downstream track-segments is prone to charge mismeasurements. We suppress such candidates with stringent requirements on a dedicated discriminant based on many low-level variables associated with track reconstruction~\cite{LHCb-PUB-2017-011}. Candidates consistent with the \Dstarp decay topology are reconstructed by computing the two-body mass \M using the known \Dz and $\pi^+$ masses~\cite{Patrignani:2016xqp} and the reconstructed momenta~\cite{Aaltonen:2011se}. The mass resolution is improved by nearly a factor of two with a kinematic fit that constrains the \Dstarp candidate to originate from a primary vertex~\cite{Hulsbergen:2005pu}. If multiple primary vertices are reconstructed, the vertex resulting from the fit with the best $\chi^2$ probability is chosen. The sample is further enriched in primary charm decays by restricting the impact-parameter chi-squared, $\chi^2_{\rm IP}$, of the \Dz and $\pi_s^+$ candidates such that the candidates point to the primary vertex. The $\chi^2_{\rm IP}$ variable is the difference between the $\chi^2$ of the primary-vertex fit reconstructed including or excluding the considered particle, and offers a measure of consistency with the hypothesis that the particle originates from the primary vertex. Only opposite-charge particle pairs with $K^\mp\pi^\pm$ mass within 24\mevcc (equivalent to approximately three times the mass resolution) of the known \Dz mass~\cite{Patrignani:2016xqp}  and $K^+K^-$ and $\pi^+\pi^-$  masses more than 40\mevcc away from the \Dz mass are retained. Accidental combinations of a genuine \Dz with a random soft pion are first suppressed by removing the 13\% of events where more than one \Dstarp candidate is reconstructed. We then use an artificial neural-network discriminant that exploits the $\pi_s^+$ pseudorapidity, transverse momentum, and particle-identification information, along with the track multiplicity of the event. The discriminant is trained on an independent RS sample to represent the WS signal features and on WS events containing multiple candidates to represent background. Finally, we remove from the WS sample events where the same \Dz candidate is also used to reconstruct a RS decay, which reduces the background by 16\% with no significant loss of signal.

\section{Yield determination}
\begin{figure}[!t]
\centering
\includegraphics[width=0.5\textwidth]{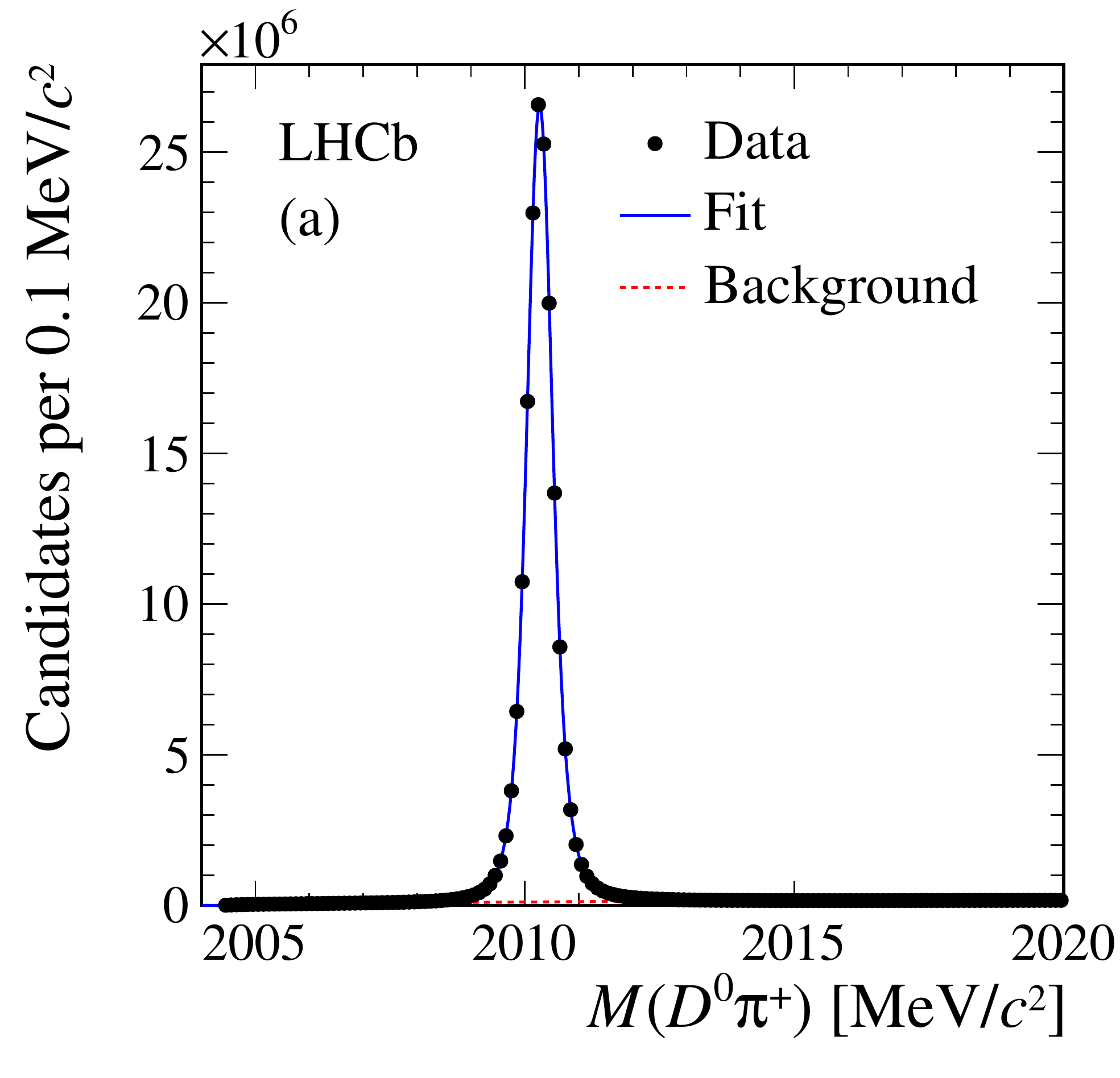}\hfil
\includegraphics[width=0.5\textwidth]{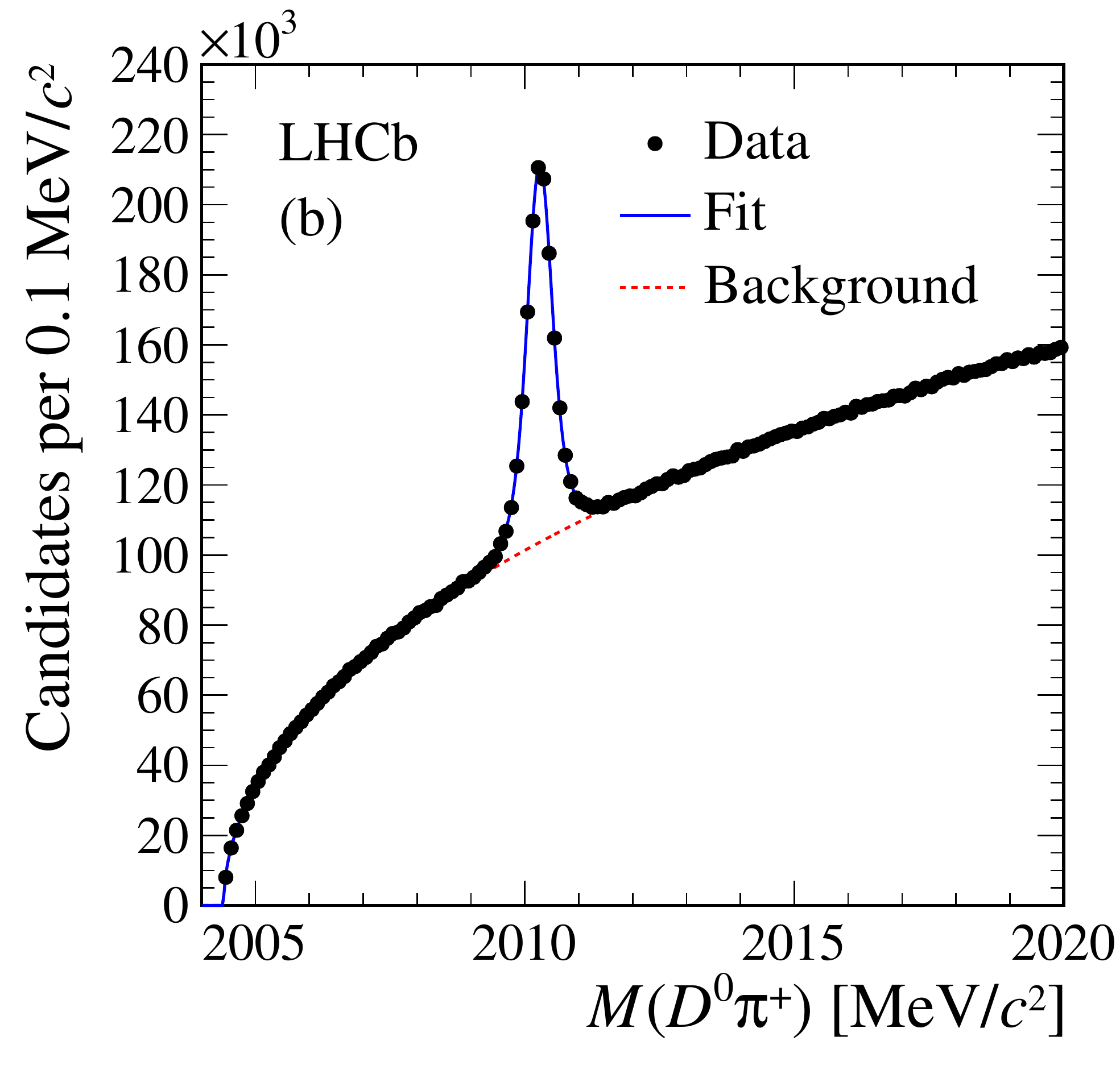}\\
\caption{\small Distribution of \M  for selected (a) right-sign $\Dz\to K^-\pi^+$ and (b) wrong-sign $\Dz\to K^+\pi^-$ candidates.\label{fig:mass}}
\end{figure}

\begin{figure}[ht]
\centering
\includegraphics[width=0.6\textwidth]{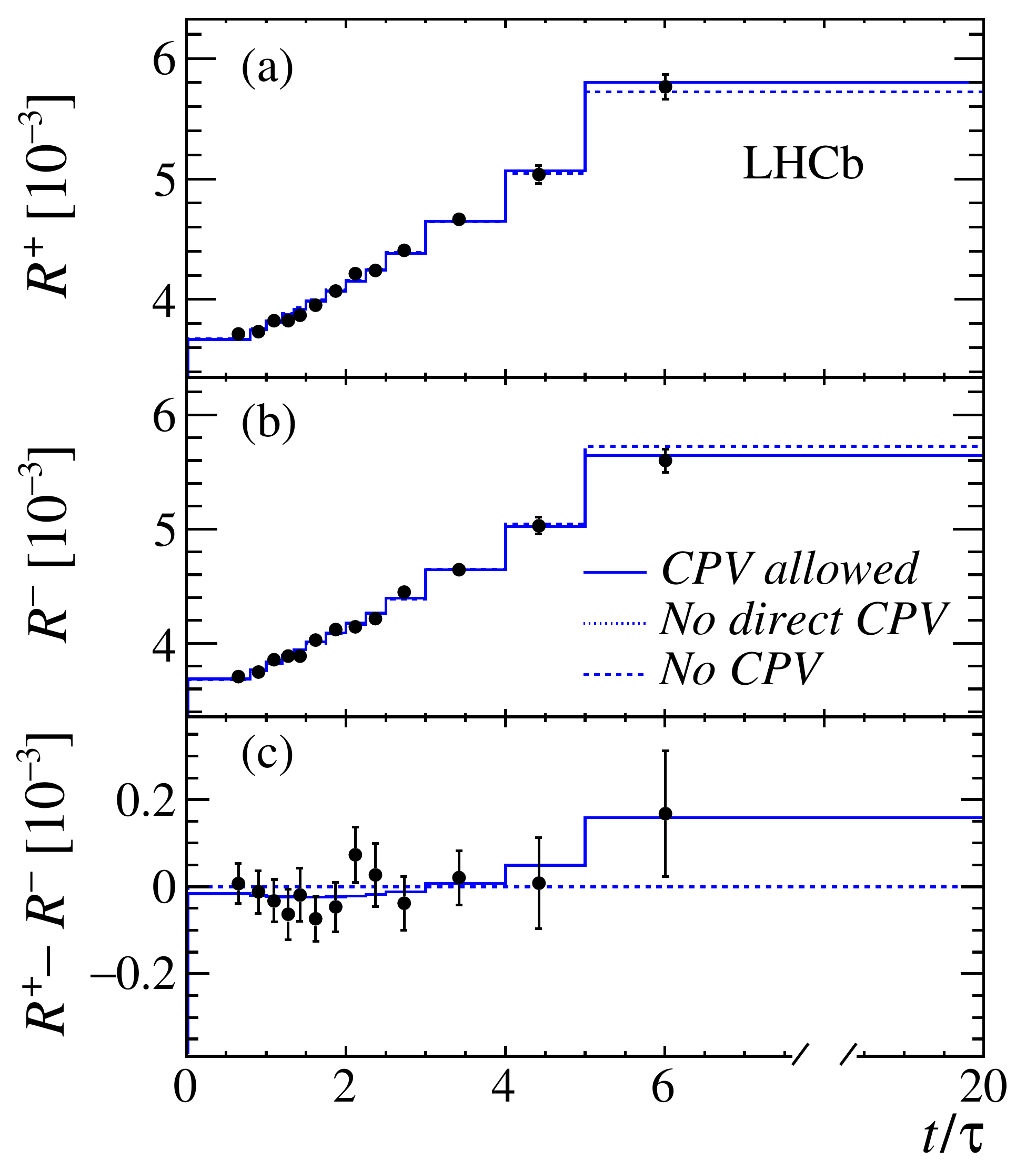}\\
\caption{\small Efficiency-corrected ratios of WS-to-RS yields for (a) \Dstarp decays, (b) \Dstarm decays, and (c) their differences as functions of decay time in units of \Dz lifetime. Projections of fits allowing for (dashed line) no \CP violation, (dotted line) no direct \CP violation, and (solid line) direct and indirect \CP violation are overlaid. The last two curves overlap. The abscissa of each data point corresponds to the average decay time over the bin. The error bars indicate the statistical uncertainties.\label{fig:finalResults}}
\end{figure}

The RS and WS signal yields are determined by fitting the \M distribution of \Dz signal candidates. The decay-time-integrated \M distributions of the selected RS and WS candidates are shown in Fig.~\ref{fig:mass}. The smooth background is dominated by favored $\Dz\to K^-\pi^+$ and $\Dzb\to K^+\pi^-$ decays associated with random soft-pion candidates. The sample contains approximately $1.77\times10^{8}$ RS and $7.22\times10^5$ WS signal decays. Each sample is divided into 13 subsamples according to the decay time, and signal yields are determined for each subsample using an empirical shape~\cite{Aaij:2012nva}. We assume that the signal shapes are common to WS and RS decays for a given \Dstar meson flavor whereas the descriptions of the backgrounds are independent. The decay-time-dependent WS-to-RS rate ratios $R^+$ and $R^-$ observed in the \Dz and \Dzb samples, respectively, and their difference, are shown in Fig.~\ref{fig:finalResults}. The ratios and difference include corrections for the relative efficiencies for reconstructing $K^-\pi^+$ and $K^+\pi^-$ final states.

\section{Determination of oscillation parameters}
The mixing parameters are determined by minimizing a $\chi^2$ function that includes terms for the difference between the observed and predicted ratios and for systematic effects,
\chisquaredexpression
The observed WS-to-RS yield ratio and its statistical uncertainty in the decay-time bin $i$ are denoted by $r_i^{\pm}$ and $\sigma_i^{\pm}$, respectively. The associated predicted value $\Ripred^{\pm}$ corresponds to the decay-time integral over bin $i$ of Eq.~\eqref{eq:true-ratio}, including bin-specific corrections. The parameters associated with  these corrections are determined separately for data collected in different LHC and detector configurations and vary independently in the fit within their constraint $\chi^2_{\rm corr}$ in Eq.~\eqref{eqn:fit}. Such corrections account for small biases due to (i) the decay-time evolution of the 1\%--10\% fraction of signal candidates originating from \bquark-hadron decays, (ii) the approximately $0.3\%$ component of the background from misreconstructed charm decays that peak in the signal region, and (iii) the effect of instrumental asymmetries in the $K^{\pm}\pi^{\mp}$ reconstruction efficiencies.  
The secondary-\Dstarp fraction is determined by fitting, in each decay-time bin, the  $\chi^2_{\rm IP}$ distribution of RS \Dz signal decays.  The peaking background, dominated by $\Dz \to K^-\pi^+$ decays in which both final-state particles are misidentified, is determined by extrapolating into the \Dz signal mass region the contributions from misreconstructed charm decays identified by reconstructing the two-body mass under various mass hypotheses for the decay products. The relative efficiency $\epsilon_r^\pm$ accounts for the effects of instrumental asymmetries in the $K^{\pm}\pi^{\mp}$ reconstruction efficiencies, mainly caused by \Km mesons having a larger nuclear interaction cross-section with matter than \Kp mesons. These asymmetries are measured in data to be typically 0.01 with 0.001 precision, independent of decay time. They are derived from the efficiency ratio $\epsilon_r^+=1/\epsilon_r^-=\epsilon(K^+\pi^-)/\epsilon(K^-\pi^+)$, obtained by comparing the ratio of $D^- \to K^+\pi^-\pi^-$ and $D^- \to \KS (\to \pi^+\pi^-)\pi^-$ yields with the ratio of the corresponding charge-conjugate decay yields. The asymmetry between \Dp and \Dm production rates~\cite{LHCb:2012fb} cancels in this ratio, provided that the kinematic distributions are consistent across samples. We therefore weight the $D^- \to K^+ \pi^- \pi^-$  candidates so that their kinematic distributions match those in the $D^- \to \KS \pi^-$  sample. We then determine $\epsilon_r^\pm$ as functions of kaon momentum to account for the known momentum-dependence of the asymmetry between $K^+$ and $K^-$ interaction rates with matter. In addition, a systematic uncertainty for possible residual contamination from spurious soft pions is included through a 1.05--1.35 scaling of the overall uncertainties. The scaling value is chosen such that a fit with a constant function of the time-integrated WS-to-RS ratio versus false-pion probability has unit reduced $\chi^2$.\par
The observed WS-to-RS yield ratios for the \Dz and \Dzb samples are studied first with bin-by-bin arbitrary offsets designed to mimic the effect of significantly different mixing parameters in the two samples. To search for residual systematic uncertainties, the analysis is repeated on statistically independent data subsets chosen according to criteria likely to reveal biases from specific instrumental effects. These criteria include the data-taking year (2011--2012 or 2015--2016), the magnet field orientation, the number of primary vertices in the event, the candidate multiplicity per event, the trigger category, the \Dz momentum and $\chi^2_{\rm IP}$ with respect to the primary vertex, and the per-candidate probability to reconstruct a spurious soft pion. The resulting variations of the measured \CP-averaged and \CP-violating parameters are consistent with statistical fluctuations, with  $p$-values distributed uniformly in the 4\%--85\% range.

\begin{table}[hb!!]
\centering
\caption{\small Results of fits for different \CP-violation hypotheses. The first contribution to the uncertainties is statistical and the second systematic. Correlations include both statistical and systematic contributions. \label{tab:finalResults}}
\resizebox{\textwidth}{!}{
\begin{tabular}{lr@{\,$\pm$\,}c@{\,$\pm$\,}lrrrrrr}
\hline\hline
\multicolumn{4}{c}{Results [$10^{-3}$] } & \multicolumn{6}{c}{Correlations} \\
\hline
\multicolumn{10}{c}{Direct and indirect \CP violation}\\
Parameter & \multicolumn{3}{c}{Value} & $R_D^+$ & $y'^+$ & $(x'^{+})^2$ & $R_D^-$ & $y'^{-}$ & $(x'^{-})^2$ \\
\hline
$R_D^+$   & $3.454$ & $0.040$ & $0.020$ & $1.000$ & $-0.935$ &  $0.843$ & $-0.012$ & $-0.003$ &  $0.002$ \\
$y'^+$    &  $5.01$ & $0.64$ & $0.38$  &         &  $1.000$ & $-0.963$ & $-0.003$ &  $0.004$ & $-0.003$ \\
$(x'^{+})^2$ & $0.061$ & $0.032$ & $0.019$ &         &          &  $1.000$ &  $0.002$ & $-0.003$ &  $0.003$ \\
$R_D^-$   & $3.454$ & $0.040$ & $0.020$ &         &          &          &  $1.000$ & $-0.935$ &  $0.846$ \\
$y'^-$    &  $5.54$ & $0.64$ & $0.38$  &         &          &          &          &  $1.000$ & $-0.964$ \\
$(x'^{-})^2$ & $0.016$ & $0.033$ & $0.020$ &         &          &          &          &          &  $1.000$ \\
\hline
\multicolumn{10}{c}{No direct \CP violation}\\
Parameter & \multicolumn{3}{c}{Value} & $R_D$ & $y'^+$ &  $(x'^{+})^2$ & $y'^-$ & $(x'^{-})^2$ \\
\hline
$R_D$     & $3.454$ & $0.028$ & $0.014$ & $1.000$ & $-0.883$ &  $0.745$ & $-0.883$ &  $0.749$ \\
$y'^+$    &  $5.01$ & $0.48$ & $0.29$  &         &  $1.000$ & $-0.944$ &  $0.758$ & $-0.644$ \\
$(x'^{+})^2$ & $0.061$ & $0.026$ & $0.016$ &         &          &  $1.000$ & $-0.642$ &  $0.545$ \\
$y'^-$    &  $5.54$ & $0.48$ & $0.29$  &         &          &          &  $1.000$ & $-0.946$ \\
$(x'^{-})^2$ & $0.016$ & $0.026$ & $0.016$ &         &          &          &          &  $1.000$ \\
\hline
\multicolumn{10}{c}{No \CP violation}\\
Parameter & \multicolumn{3}{c}{Value} & $R_D$ & $y'$ & $x'^2$ \\
\hline
$R_D$  & $3.454$ & $0.028$ & $0.014$ & $1.000$ & $-0.942$ &  $0.850$ \\
$y'$   &  $5.28$ & $0.45$ & $0.27$  &         &  $1.000$ & $-0.963$ \\
$x'^2$ & $0.039$ & $0.023$ & $0.014$ &         &          &  $1.000$ \\
\hline\hline
\end{tabular}}
\end{table}

\section{Results}
\begin{figure*}[t]
\centering
\includegraphics[width=\textwidth]{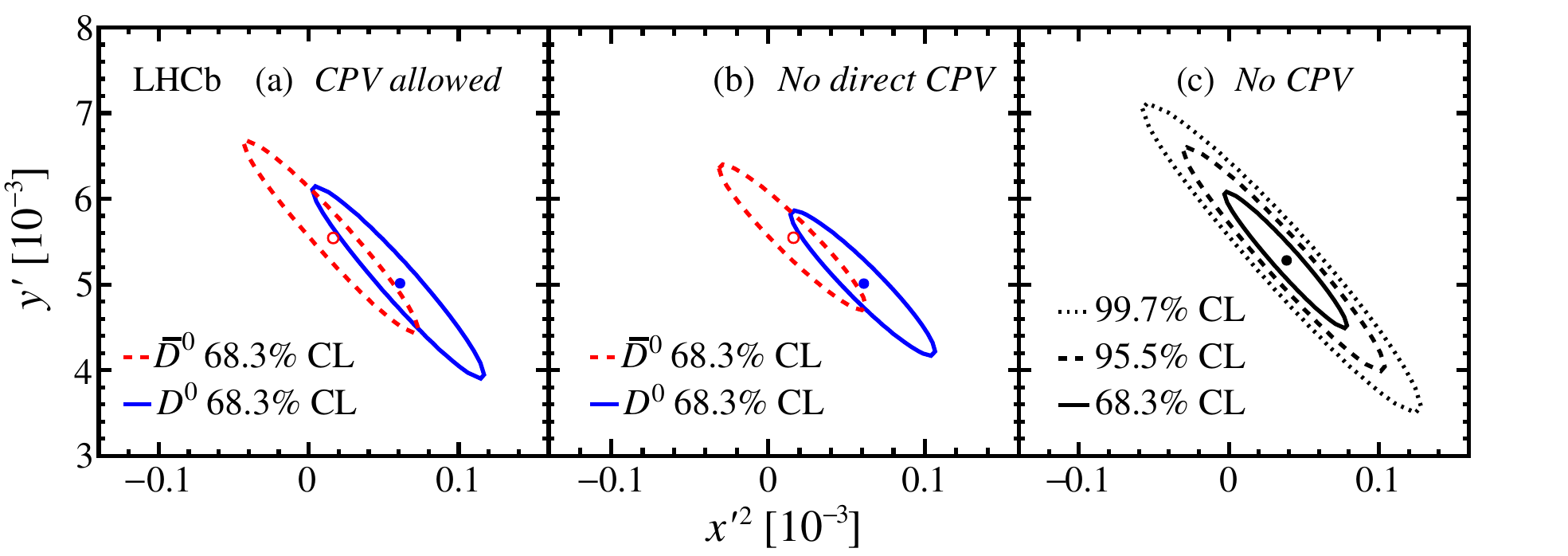}\\
\caption{Two-dimensional confidence regions in the $(x'^2,y')$ plane obtained (a) without any restriction on \CP violation, (b) assuming no direct \CP violation,  and (c) assuming \CP conservation. The dashed (solid) curves in (a) and (b) indicate the contours of the mixing parameters associated with \Dzb (\Dz) decays. The best-fit value for \Dzb (\Dz) decays is shown with an open (filled) point. The solid, dashed, and dotted curves in (c) indicate the contours of \CP-averaged mixing parameters at 68.3\%, 95.5\%, and 99.7\% confidence levels (CL), respectively, and the point indicates the best-fit value.
\label{fig:contours}}
\end{figure*}

The efficiency-corrected WS-to-RS yield ratios are subjected to three fits. The first fit allows for direct and indirect \CP violation; the second allows only for indirect \CP violation by imposing $R_D^+ = R_D^-$; and the third is a fit under the \CP-conservation hypothesis,  in which all mixing parameters are common to the \Dz and \Dzb samples. The fit results and their projections are presented in Table~\ref{tab:finalResults} and Fig.~\ref{fig:finalResults}, respectively. Figure~\ref{fig:contours} shows the central values and confidence regions in the $(x'^2,\, y')$ plane. For each fit, 208 WS-to-RS ratio data points are used, corresponding to 13 ranges of decay time; distinguishing \Dstarp from \Dstarm decays; two magnetic-field orientations; and 2011, 2012, 2015, and 2016 data sets. The consistency of the data with the hypothesis of \CP symmetry is determined from the change in $\chi^2$ probability between the fit that assumes \CP conservation and the fit in which \CP violation is allowed. The resulting $p$-value is 0.57 (0.37) for the fit in which both direct and indirect (indirect only) \CP violation is allowed, showing that the data are compatible with \CP symmetry.

The fit uncertainties incorporate both statistical and systematic contributions. The statistical uncertainty, determined in a separate fit by fixing all nuisance parameters to their central values, dominates the total uncertainty.  The systematic component is obtained by subtraction in quadrature. The leading systematic uncertainty is due to residual secondary-\Dstarp contamination and does not exceed half of the statistical uncertainty. The second largest contribution is due to spurious soft pions. Smaller effects are due to peaking backgrounds for the \CP-averaged results, and uncertainties in detector asymmetries  for the  \CP-violating results. All reported results, $p$-values, and the contours shown in Fig.~\ref{fig:contours}, include total uncertainties. 

Direct \CP violation would produce a nonzero intercept at $t=0$ in the efficiency-corrected difference of WS-to-RS yield ratios between \Dz and \Dzb mesons shown in Fig.~\ref{fig:finalResults}\,(c). We  parametrize this effect with the asymmetry measured in the fit that allows for direct \CP violation, $A_D~\equiv~{ (R_D^+ - R_D^-) / (R_D^+ + R_D^-) }~=~(-0.1\pm8.1\pm4.2)\times10^{-3}$, where the first uncertainty is statistical and the second systematic. Indirect \CP violation would result in a time dependence of the efficiency-corrected difference of yield ratios, which is not observed in Fig.~\ref{fig:finalResults}\,(c). From the results of the fit allowing for direct and indirect \CP violation, a likelihood for $|q/p|$ is constructed using the relations $x'^\pm = |q/p|^{\pm 1}(x'\cos\phi \pm y'\sin\phi)$ and $y'^\pm = |q/p|^{\pm 1}(y'\cos\phi \mp x'\sin\phi)$. Confidence intervals are derived with a likelihood-ratio ordering~\cite{Feldman:1997qc}, assuming that the parameter correlations are independent of the true values of the mixing parameters. We determine $1.00 < |q/p| < 1.35$ and $0.82 < |q/p| < 1.45$ at the $68.3\%$ and $95.5\%$ confidence levels, respectively.

The $R_D$ result departs from the previous result based on a subset of the same data~\cite{Aaij:2013wda}, which was biased by the then-undetected residual spurious-pion background. Since such background induces an apparent global shift toward higher WS-to-RS ratio values, the bias affects predominantly the $R_D$ measurement and less severely the mixing-parameter determination. The systematic uncertainties are significantly reduced because the dominant components are statistical in nature or sensitive to a generally improved understanding of the data quality.

\section{Summary}
We study \Dz--\Dzb oscillations using $\Dstarp\to \Dz (\to K^+\pi^-) \pi^+$ decays reconstructed in a data sample of $pp$ collisions collected by the \lhcb experiment from 2011 through 2016, corresponding to an integrated luminosity of $5.0\invfb$. Assuming \CP conservation, the mixing parameters are measured to be $x'^2=(3.9 \pm 2.7) \times10^{-5}$, $y'= (5.28 \pm 0.52) \times 10^{-3}$, and $R_D = (3.454 \pm 0.031)\times10^{-3}$. The results are twice as precise as previous LHCb results~\cite{Aaij:2013wda} that were based on a subset of the present data, and supersede them. Studying \Dz and \Dzb decays separately shows no evidence for \CP violation and provides the current most stringent bounds on the parameters $A_D$ and $|q/p|$ from a single measurement, $A_D =(-0.1\pm9.1)\times10^{-3}$ and $1.00< |q/p| <1.35$ at the $68.3\%$ confidence level.

\section*{Acknowledgements}
%
%
\noindent We express our gratitude to our colleagues in the CERN
accelerator departments for the excellent performance of the LHC. We
thank the technical and administrative staff at the LHCb
institutes. We acknowledge support from CERN and from the national
agencies: CAPES, CNPq, FAPERJ and FINEP (Brazil); MOST and NSFC
(China); CNRS/IN2P3 (France); BMBF, DFG and MPG (Germany); INFN
(Italy); NWO (The Netherlands); MNiSW and NCN (Poland); MEN/IFA
(Romania); MinES and FASO (Russia); MinECo (Spain); SNSF and SER
(Switzerland); NASU (Ukraine); STFC (United Kingdom); NSF (USA).  We
acknowledge the computing resources that are provided by CERN, IN2P3
(France), KIT and DESY (Germany), INFN (Italy), SURF (The
Netherlands), PIC (Spain), GridPP (United Kingdom), RRCKI and Yandex
LLC (Russia), CSCS (Switzerland), IFIN-HH (Romania), CBPF (Brazil),
PL-GRID (Poland) and OSC (USA). We are indebted to the communities
behind the multiple open-source software packages on which we depend.
Individual groups or members have received support from AvH Foundation
(Germany), EPLANET, Marie Sk\l{}odowska-Curie Actions and ERC
(European Union), ANR, Labex P2IO, ENIGMASS and OCEVU, and R\'{e}gion
Auvergne-Rh\^{o}ne-Alpes (France), RFBR and Yandex LLC (Russia), GVA,
XuntaGal and GENCAT (Spain), Herchel Smith Fund, the Royal Society,
the English-Speaking Union and the Leverhulme Trust (United Kingdom).

\addcontentsline{toc}{section}{References}
\setboolean{inbibliography}{true}
\bibliographystyle{LHCb}
\bibliography{main}

 
\cleardoublepage
\centerline{\large\bf LHCb collaboration}
\begin{flushleft}
\small
R.~Aaij$^{40}$,
B.~Adeva$^{39}$,
M.~Adinolfi$^{48}$,
Z.~Ajaltouni$^{5}$,
S.~Akar$^{59}$,
J.~Albrecht$^{10}$,
F.~Alessio$^{40}$,
M.~Alexander$^{53}$,
A.~Alfonso~Albero$^{38}$,
S.~Ali$^{43}$
% ...

{\footnotesize \it
% ...
$ ^{5}$Clermont Universit{\'e}, Universit{\'e} Blaise Pascal, CNRS/IN2P3, LPC, Clermont-Ferrand, France\\
% ...
$ ^{10}$Fakult{\"a}t Physik, Technische Universit{\"a}t Dortmund, Dortmund, Germany\\
% ...
$ ^{38}$ICCUB, Universitat de Barcelona, Barcelona, Spain\\
$ ^{39}$Instituto Galego de F{\'\i}sica de Altas Enerx{\'\i}as (IGFAE), Universidade de Santiago de Compostela, Santiago de Compostela, Spain\\
$ ^{40}$European Organization for Nuclear Research (CERN), Geneva, Switzerland\\
% ...
$ ^{43}$Nikhef National Institute for Subatomic Physics, Amsterdam, The Netherlands\\
% ...
$ ^{48}$H.H. Wills Physics Laboratory, University of Bristol, Bristol, United Kingdom\\
% ...
$ ^{53}$School of Physics and Astronomy, University of Glasgow, Glasgow, United Kingdom\\
% ...
$ ^{59}$University of Cincinnati, Cincinnati, OH, United States\\
% ...
}
\end{flushleft}

\end{document}